# Remote-excitation and remote-detection of single quantum dot using propagating surface plasmons on silver nanowire*

LI Qiang[1](李强), WEI Hong[1](魏红)**, XU Hong-Xing[1,2](徐红星)

[1] *Institute of Physics, Chinese Academy of Sciences, and Beijing National Laboratory for Condensed Matter Physics, Beijing 100190, China*

[2] *Center for Nanoscience and Nanotechnology, and School of Physics and Technology, Wuhan University, Wuhan 430072, China*

* Project supported by the National Natural Science Foundation of China (Grant Nos. 11374012, 11134013, 11227407 and 61210017), the Ministry of Science and Technology of China (Grant No. 2012YQ12006005), and the Knowledge Innovation Project of Chinese Academy of Sciences (Grant No. KJCX2-EW-W04).

**Corresponding author. E-mail: weihong@iphy.ac.cn

Using propagating surface plasmons (SPs) on silver nanowire (NW), we demonstrate that focused laser light at the end of silver nanowire can excite single quantum dot (QD) microns away from the excitation spot. The QD-NW interaction allows the excited QD convert part of its energy into propagating SPs which then can be detected at the remote sites. Simultaneous multi-QDs remote-excitation and detection are also realized. Furthermore, the tight confinement of propagating SPs around the NW surface enables selective excitation of QDs very close in space, which cannot be realized under conventional excitation condition. This remote excitation and detection approach may find applications in optical imaging and sensing of chemical and biological systems.

## 1. Introduction

Colloidal semiconductor quantum dots (QDs) with a diameter of a few nanometers have many attractive properties such as size tunable emission, wide band absorption, high photoluminescence quantum yield and exceptional photostability against photobleaching.[1, 2] These properties make QDs promising for many applications such as light-emitting diodes,[3] single-molecule tracking,[4] single-photon sources[5] and so on. In conventional QD photoluminescence measurements, the incident laser light illuminates the QDs directly and the fluorescence is collected from the same area. Although such an excitation and detection approach can be easily performed and is widely used, it may not be feasible in some applications, especially in living systems, where the higher power incident light might cause cell destruction or induce chemical modifications of the analytes. Recently, a novel technique for performing surface-enhanced Raman scattering (SERS) using propagating surface plasmons (SPs) on silver nanowire (NW) as a remote excitation source rather than direct optical excitation has been reported.[6] This approach allows remote-excitation SERS sensing and has great potential to expand ultrasensitive chemical detection to new systems.

Due to the excitation of surface plasmons, collective oscillations of conduction electrons at metal surface, metal (especially gold and silver) nanostructures show extraordinary optical properties, for instance, large electromagnetic field enhancement,[7, 8] tunable and sensitive SP resonances,[9-11] and light propagation with subwavelength field confinement. The subwavelength light guiding effect is of enormous interest due to its potential applications in building nanophotonic circuits, as well as integrating photonic and electronic circuits to overcome the limitations of bandwidth and data transmission rates of electrical circuits.[12] Metal nanowires are important elements supporting propagating SPs and have attracted much attention in recent years.[13-15] The propagating SPs on metal NWs can be excited by focusing incident light onto symmetry-broken positions, such as the end or a sharp corner of the NW and the interconnecting junctions in NW networks, and the SPs can also be converted to free-space photons at those positions.[16-19] Based on the propagating SPs on metal NWs, signal processing functionalities including routing,[20, 21] spectral splitting,[22] logic functions,[23, 24] and modulation [25] have been demonstrated. The field confinement and propagation of SPs also make metal NWs attractive for investigating light-matter interactions in NW-emitter coupled systems.[17, 26-28]

In this paper, we demonstrate that the propagating SPs on silver NW can be used to realize remote excitation and remote detection of single QD fluorescence. The QD located near the NW surface can be remotely excited by the propagating SPs generated by the focused laser light at the end of the silver NW. The exciton-plasmon coupling makes the QD convert part of its energy into SPs propagating along the NW which finally scattered as photons at the NW ends. This new excitation and detection approach avoids direct illumination on the area for measurements, and may find applications in photosensitive systems in chemical and biological studies.

## 2. Experimental method

In experiments, silver NWs were synthesized using a solution-phase polyol method.[29] The diameter of the NWs used in this study is about 80 nm. The Ag NWs in a droplet of ethanol were deposited onto glass slides and dried naturally. The slides with Ag NWs were immediately covered by $Al_2O_3$ film of 10 nm thickness using atomic layer deposition (ALD) system (Cambridge NanoTech; Savannah-100) operating at 200 ℃. Then highly luminescent organic CdSe/ZnS QDs (Qdot® 655 ITK™, Invitrogen) were spin coated onto the sample. Optical measurements were carried out using an inverted microscope (IX71, Olympus). Laser light of 532 nm wavelength was focused onto the sample using a 100× oil immersion objective (NA 1.4, Olympus). The laser spot size on sample was about 0.6 μm. The fluorescence from the QDs was collected by the same objective and detected by an EMCCD camera (ixon DV887, Andor) which records the time traces of the fluorescence counts with exposure time of 100 ms for each frame. The polarization of the laser light was adjusted by rotating a half-wave plate. The laser power is about 4.0 μW for QD direct excitation and 60 μW for the remote excitation.

## 3. Results and discussion

Figure 1 shows the experimental results of a QD-NW system under conventional excitation condition, i.e. the laser light is directly focused on the QD (Fig. 1(a)). Figure 1(b) and 1(c) show the optical transmission image and corresponding fluorescence image of a NW-QD structure. After excited by the focused laser light, the QD emits photons into the free space which are detected as the large bright spot marked as A. Meanwhile, energy from the QD can also convert to propagating SPs on the NW and finally scatter out as photons at the wire ends B and C (shown by the two smaller bright spots).[17, 28] Time traces of fluorescence counts from the coupled QD A and scattered photons at B and C (Fig. 1d) show blinking behavior, that is, the emission is randomly switched between ON (bright) and OFF (dark) states under continuous excitation. The blinking phenomenon indicates only one QD is excited.[30, 31] The high degree of correlation between the time trace of the fluorescence counts from the QD (top curve in Fig. 1d) and those from the ends of the NW (the second and third curves in Fig. 1d) validates the exciton-plasmon conversion. This result also indicates that the QD fluorescence signal can be remotely detected at the wire ends, micrometers away from the QD position.

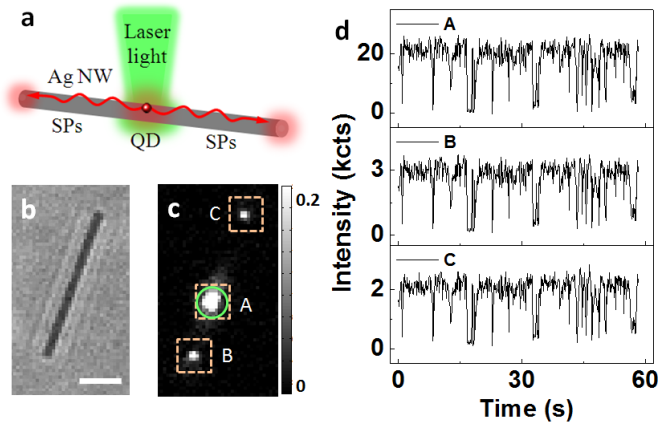

Fig. 1. (a) Sketch of the Ag NW-QD system. The QD is directly excited by 532 nm laser light. (b) Optical transmission image of a Ag NW. The scale bar is 2 μm. (c) Fluorescence image showing the coupling of single QD to the Ag NW. The green circle shows the laser excitation position. The largest bright spot A corresponds to the fluorescence from the QD, while two smaller spots B and C correspond to SPs scattered from the two ends of the NW. (d) Time traces of fluorescence counts of QD A and scattered light at the NW end B and C. The intensity unit kcts means 1000 counts. The light pink boxes in (c) show the regions where the counts of each pixel are integrated to generate the emission counts.

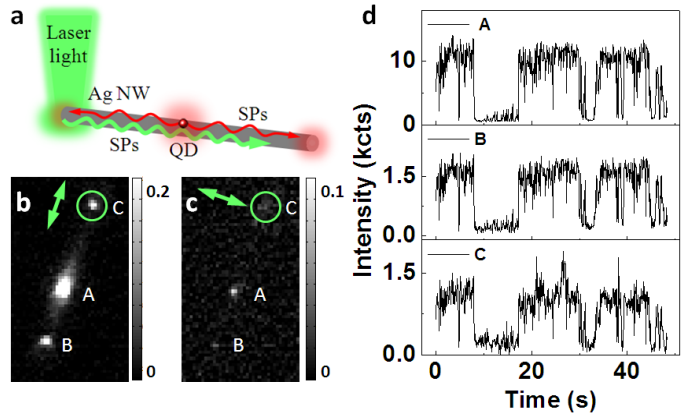

**Fig. 2.** (a) Sketch for remote excitation of single QD using propagating SPs (denoted by green wavy line) generated by focused light at the end of the NW. (b, c) Fluorescence images showing the remote excitation of QD A with laser polarization parallel (b) and perpendicular (c) to the NW. The green circles show the laser excitation positions. (d) Time traces of fluorescence counts of QD A and scattered light at B and C with the laser polarized along the NW.

Then we moved the wire end C to the excitation spot to excite the propagating SPs as schematically shown in Fig. 2(a). When the laser light was polarized along the NW, strong fluorescence signals from the QD A were obtained (see Fig. 2(b)). In order to verify that the QD A is excited by the propagating SPs generated at the wire end C, we changed the polarization of laser light to be perpendicular to the NW. In this case, the QD emission intensity is strongly decreased (see Fig. 2(c)), which is caused by the lower generation efficiency of propagating SPs when the polarization is perpendicular to the NW.[32] Thus the emission of the QD is a result of propagating SPs excitation. Except the bright emission spot at A, both wire the NW ends lit up. From the high degree of correlation between the time traces of the fluorescence counts measured at A, B and C (Fig. 2d), we can conclude that the wire end emission is from the propagating SPs generated by the excited QD. The unrelated component in signal from terminal C (bottom curve in Fig. 2d) is mainly from the emission of alumina-coated nanowire and substrate under laser excitation. These results demonstrate that, mediated by the propagating SPs on the NW, we can realize both the remote excitation and the remote detection of QD fluorescence signal.

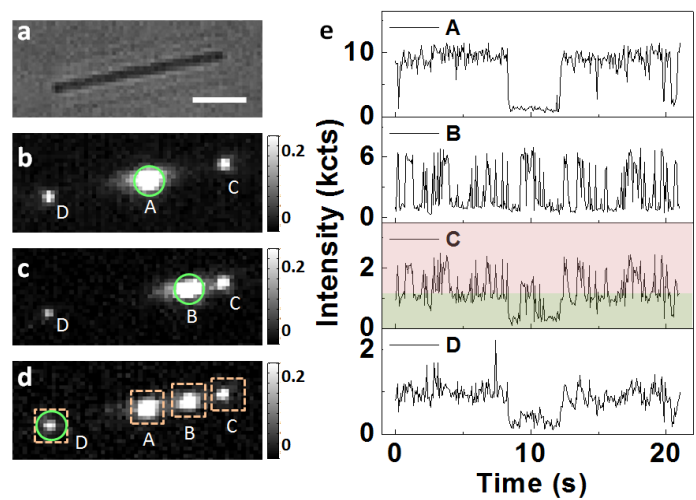

**Fig. 3.** (a) Optical transmission image of a Ag NW. The scale bar is 2 μm. (b, c) Fluorescence images showing the coupling of QD A and B to the Ag NW. The green circles show the laser excitation positions. The bright spots A and B correspond to the QD fluorescence, while two smaller spots C and D correspond to SPs scattered from the NW ends. (d) Fluorescence images showing the simultaneous excitation of a pair of QDs excited by propagating SPs which are generated by focusing laser light at NW end D. (e) Time traces of fluorescence counts of QDs A and B, and scattered light at the NW ends C and D. The pink area and the cyan area are used to separate the two bright states.

Furthermore, simultaneous multi-QDs remote-excitation and detection can also be achieved. Figure 3(a) is an optical image of a silver NW. It happens that there are two QDs coupled with this NW. When the laser light was focused onto each QD, both wire ends C and D show photon emissions (Fig. 3(b) and 3(c)). Then we moved the NW terminal D to the excitation spot and the laser polarization was along the NW. It can be seen from Fig. 3(d) that both QDs are excited by the propagating SPs and the wire ends emit photons converted from QD-generated SPs. Time traces of fluorescence counts from the two QDs and the wire ends are shown in Fig. 3(e). Clearly, the time trace curves of both QD A and B show binary blinking behavior. The difference in the blinking behavior of QD A and B might be from the difference in the surface structure of the two QDs.[33, 34] The blinking curves of the scattered photons at C show two-level ON states (see the third curve in Fig. 3e), indicating that both QD A and B were efficiently coupled with the NW and they both contributed to the generation of propagating SPs. Because of the longer distance between D and B, little energy component from the QD B is observed at the wire end D (see the bottom curve in Fig. 3e).

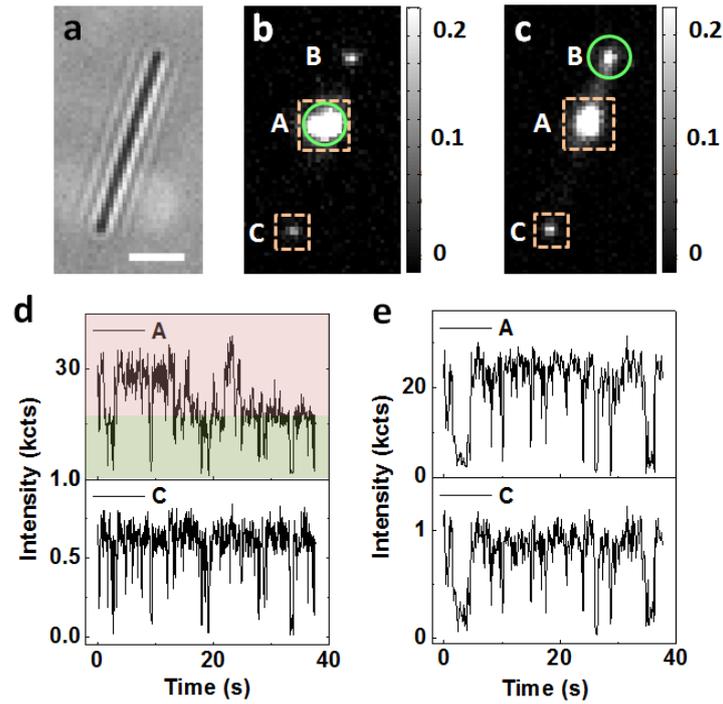

**Fig. 4.** (a) Optical transmission image of the Ag NW. A pair of QDs is distributed in the proximity of the Ag NW. The scale bar is 2 μm. (b, d) Fluorescence images and time traces of fluorescence counts showing the coupling of QD pair with the Ag NW under QD direct excitation conditions. The light pink boxes show the integration regions used to calculate the emitting counts. The time traces of fluorescence counts from the QD pair shows two-level ON states (separated by the pink area and cyan area), indicating that both QDs are excited. (c, e) Fluorescence image and time traces of fluorescence counts under wire end B excitation condition. The light green circles show the laser excitation positions.

As the propagating SPs on the NW is highly confined around the NW surface, only the QD located at the near field area of the NW can be excited. Thus, we can selectively excite one of two QDs close to each other and that usually cannot be achieved under conventional excitation conditions. Figure 4 shows a system composed of a Ag NW and two QDs. When the laser light was directly focused on the QDs, the blinking curve of the QD emission (bright spot A in Fig. 4b) shows two-level ON states (see the top curve in Fig. 4d), indicating that both QDs are excited. It is also found that the emission counts at C show only one ON state (see the bottom curve in Fig. 4d) and is partly correlated with the time trace at A, indicating only one of the two QDs is coupled with the NW. Then we moved the NW end B to the excitation spot to remotely excite the QDs using propagating SPs (Fig. 4(c)). The single-ON-state and highly correlated blinking behavior of the emissions at spots A, B and C (Fig. 4(e)) indicates that only one QD is excited by the propagating SPs, and the excited QD converted parts of its energy to the propagating SPs on the NW and finally reemitted as photons at wire ends B and C. The tight spatial confinement of the propagating SPs on the NW guarantees that only the QD near the NW is selectively excited.

Compared with the fluorescence intensity of QDs, the scattered light at the wire ends is weak. This is mainly determined by two factors. Firstly, the SP-generation efficiency, defined as the probability of QD energy converted into propagating SPs, is strongly dependent on the QD-NW separation.[28] If the separation is too large, the SP-generation efficiency is quite small. While if the separation is too small, most of the QD energy is dissipated non-radiatively.[35] Thus the QD-NW separation need be optimized to achieve the highest SP-generation efficiency. Secondly, the Ohmic loss of propagating SPs on silver NW results in the lower emission intensity at the wire ends and limits the long-distance transmission of fluorescence signals from the QDs. A possible approach to solve this problem is to integrate silver NW with a dielectric nanofiber.[36] By coupling the SPs on the NW into the nanofiber, the propagation loss can be reduced.

## 4. Conclusion

In summary, we demonstrate a novel technique for performing remote excitation and remote detection of single QD using propagating SPs on the NW. Because of the interaction between propagating SPs on silver NW and excitons generated in QDs, the propagating SPs launched at the wire end can excite QDs microns away from the excitation spot, which enables the remote-excitation of QDs. In the reverse process, the excited QDs can generate SPs propagating along the NW and finally emitted as photons at the NW ends microns away from the QDs, which makes the remote-detection of QD emission feasible. Our results show that both the remote-excitation and remote-detection can be achieved at single QD level, indicating the high sensitivity of this experimental scheme. As the propagating SPs along the NW are highly confined around the NW surface, only the QDs located at the near field area of the NW can be excited, which makes the remote excitation and detection technique useful for optical imaging and sensing in chemical and biological systems with high spatial resolution and low background noise.

**Acknowledgment**

The authors thank the Laboratory of Microfabrication in Institute of Physics, Chinese Academy of Sciences for experimental support.

**References**

(1)   Medintz I L, Uyeda H T, Goldman E R and Mattoussi H 2005 *Nat. Mater.* **4** 435
(2)   Bruchez M, Moronne M, Gin P, Weiss S and Alivisatos A P 1998 *Science* **281** 2013
(3)   Sun Q J, Wang Y A, Li L S, Wang D Y, Zhu T, Xu J, Yang C H and Li Y F 2007 *Nat. Photonics* **1** 717
(4)   Courty S, Luccardini C, Bellaiche Y, Cappello G and Dahan M 2006 *Nano Lett.* **6** 1491
(5)   Brokmann X, Messin G, Desbiolles P, Giacobino E, Dahan M and Hermier J P 2004 *New J. Phys.* **6** 099
(6)   Fang Y R, Wei H, Hao F, Nordlander P and Xu H X 2009 *Nano Lett.* **9** 2049
(7)   Xu H X, Bjerneld E J, Käll M and Börjesson L 1999 *Phys. Rev. Lett.* **83** 4357
(8)   Xu H X, Aizpurua J, Käll M and Apell P 2000 *Phys. Rev. E* **62** 4318
(9)   Wei H, Reyes. Coronado A, Nordlander P, Aizpurua J and Xu H X 2010 *Acs Nano* **4** 2649

(10) Chen L, Wei H, Chen K Q and Xu H X 2014 *Chin. Phys. B* **23** 027303
(11) Xu H X and Kall M 2002 *Sensor Actuat. B-Chem.* **87** 244
(12) Ozbay E 2006 *Science* **311** 189
(13) Wei H and Xu H X 2012 *Nanophotonics* **1** 155
(14) Guo X, Ma Y G, Wang Y P and Tong L M 2013 *Laser Photonics Rev.* **7** 855
(15) Pan D, Wei H and Xu H X 2013 *Chin. Phys. B* **22** 097305
(16) Sanders A W, Routenberg D A, Wiley B J, Xia Y N, Dufresne E R and Reed M A 2006 *Nano Lett.* **6** 1822
(17) Wei H, Ratchford D, Li X E, Xu H X and Shih C K 2009 *Nano Lett.* **9** 4168
(18) Fang Z Y, Fan L R, Lin C F, Zhang D, Meixner A J and Zhu X 2011 *Nano Lett.* **11** 1676
(19) Zhu Y, Wei H, Yang P F and Xu H X 2012 *Chin. Phys. Lett.* **29** 077302
(20) Fang Y R, Li Z P, Huang Y Z, Zhang S P, Nordlander P, Halas N J and Xu H X 2010 *Nano Lett.* **10** 1950
(21) Wei H, Zhang S P, Tian X R and Xu H X 2013 *Proc. Natl. Acad. Sci. USA* **110** 4494
(22) Hu Q, Xu D H, Zhou Y, Peng R W, Fan R H, Fang N X, Wang Q J, Huang X R and Wang M 2013 *Sci. Rep.* **3** 03095
(23) Wei H, Li Z P, Tian X P, Wang Z X, Cong F Z, Liu N, Zhang S P, Nordlander P, Halas N J and Xu H X 2011 *Nano Lett.* **11** 471
(24) Wei H, Wang Z X, Tian X R, Kall M and Xu H X 2011 *Nat. Commun.* **2** 387
(25) Li Z P, Zhang S P, Halas N J, Nordlander P and Xu H X 2011 *Small* **7** 593
(26) Liu S D, Cheng M T, Yang Z J and Wang Q Q 2008 *Opt. Lett.* **33** 851
(27) Fedutik Y, Temnov V, Schöps O, Woggon U and Artemyev M 2007 *Phys. Rev. Lett.* **99** 136802
(28) Akimov A V, Mukherjee A, Yu C L, Chang D E, Zibrov A S, Hemmer P R, Park H and Lukin M D 2007 *Nature* **450** 402
(29) Sun Y G and Xia Y N 2002 *Adv. Mater.* **14** 833
(30) Neuhauser R G, Shimizu K T, Woo W K, Empedocles S A and Bawendi M G 2000 *Phys. Rev. Lett.* **85** 3301
(31) Galland C, Ghosh Y, Steinbruck A, Sykora M, Hollingsworth J A, Klimov V I and Htoon H 2011 *Nature* **479** 203
(32) Li Z P, Bao K, Fang Y R, Huang Y Z, Nordlander P and Xu H X 2010 *Nano Lett.* **10** 1831
(33) Mahler B, Spinicelli P, Buil S, Quelin X, Hermier J P and Dubertret B 2008 *Nat. Mater.* **7** 659
(34) Zhang A D, Dong C Q, Liu H and Ren J C 2013 *J. Phys. Chem. C* **117** 24592
(35) Anger P, Bharadwaj P and Novotny L 2006 *Phys. Rev. Lett.* **96** 113002
(36) Guo X, Qiu M, Bao J M, Wiley B J, Yang Q, Zhang X N, Ma Y G, Yu H K and Tong L M 2009 *Nano Lett.* **9** 4515